\documentclass[runningheads]{llncs}
\usepackage{graphicx}
\usepackage[export]{adjustbox}

\begin{document}
\title{\textit{The Diary of Niels}: Affective engagement through tangible interaction with museum artifacts}
\titlerunning{The Diary of Niels}
%
\author{Mette Muxoll Schou \and Anders Sundnes Løvlie\orcidID{0000-0003-0484-4668} }
\authorrunning{M. M. Schou and A. S. Løvlie}

%
\institute{IT University of Copenhagen, Denmark \email{asun@itu.dk} }

\maketitle              
\begin{abstract}
This paper presents a research through design exploration using tangible interactions in order to seamlessly integrate technology in a historical house museum. The study addresses a longstanding concern in museum exhibition design that interactive technologies may distract from the artifacts on display. Through an iterative design process including user studies, a co-creation workshop with museum staff and several prototypes, we developed an interactive installation called \textit{The Diary of Niels} that combines physical objects, RFID sensors and an elaborate fiction in order to facilitate increased visitor engagement. Insights from the research process and user tests indicate that the integration of technology and artifacts is meaningful and engaging for users, and helps introduce museum visitors to the historic theme of the exhibition and the meaning of the artifacts. The study also points to continued challenges in integrating such hybrid experiences fully with the rest of the exhibition. 

\end{abstract}

\section{Introduction}
In recent years, there has been increasing interest in the role of emotion and affect in the design of visitor experiences for museums \cite{Boehner_sengers_2005,Gregory2007-vi,Smith2018-ge}. However, along with the increasing use of digital technology to facilitate engaging visitor experiences in museums, many have voiced concern that such use of technology may be a detriment and distraction as much as a benefit \cite{vom_lehn,Rudloff2013-oi,wessel_potentials_2007,woodruffelectronic2001}. Such concerns from museum professionals often focus on the risk that visitors' attention may be drawn away from the museum artifacts \cite{Back2018-oo}. Concepts such as hybrid design and tangible interactions are offered in response \cite{Bannon2005-mj,Ciolfi2011-cw,Claisse2016-jb,Hornecker2006-zn,lovlie_gift_2019,lovlie_designing,marshall_tangible_mesch2016,pedersendesigning2019,risseeuw_authoring_2016,zancanarorecipes2015}. A recent user study indicates that museum visitors prefer tangible interaction formats over smartphone apps \cite{Petrelli2018-bd}.

This paper presents a research through design exploration of affective design with tangible interactions in a historical house museum: Greve Museum in Denmark. Historical museums face a difficult balancing act as they need to facilitate engaging visitor experiences, while also respecting their commitment to historical accuracy and authentic preservation and presentation of artifacts \cite{Gregory2007-vi}. Furthermore, house museums have a particular set of challenges, such as the fact that not just the objects on display but also the house itself is considered an historic artifact, meaning that "content and container are one" \cite{Claisse2018-vc}. Since the historical authenticity of the house is at the heart of the museum's identity, when introducing technology it is considered important to "maintain the spirit of the house" and to design for seamless experiences \cite{Claisse2018-vc}.

We explore the following research question: \textit{How can we design for affective engagement through tangible interactions with museum artifacts, while accommodating the museum's need to communicate historical and cultural knowledge?} The study contributes with a case that is comparable to the one discussed in \cite{Claisse2018-vc}, but with some key differences in addition to country and local context. In particular, the museum's requirements regarding historic accuracy demanded a design in which the technology was integrated seamlessly into the house, centering the interaction entirely on original historic artifacts, in a narrative presenting a historic, rather than fictional character. 

\section{Background}

Affect has been discussed in fields such as art history \cite{Prown1980-pk} and literature \cite{Meskin2003-rq} as well as design \cite{Boehner_depaula_2005}. In recent years, there has also been increasing interest in the role of affect in the design of museum experiences \cite{Smith2018-ge,Ryding_Fritsch}. This development is sometimes seen as part of a broader "affective turn" \cite{Blackman2012-hc} in social sciences and HCI. For Gregory and Witcomb affect is "an important means to achieve audience participation in the process of making meaning" \cite[p.~263]{Gregory2007-vi}. Witcomb sees affect as an element in developing a critical pedagogy for history museums, and has suggested that the traditional museum concept of a "pedagogy of walking" be replaced with a "pedagogy of feeling" \cite{witcomb2013,Witcomb2014-wq}. 

History museums pose challenges for experience design, as the need to facilitate audience engagement must be balanced against the museum's mission to exhibit authentic historical artifacts and knowledge. In museum studies, there is a long-standing concern that using mobile devices for interpretive artwork information will lead users to focus only on their mobile screens rather than the exhibited artifacts in front of them (cf. \cite{vom_lehn,wessel_potentials_2007}). In the words of Woodruff et al. \cite{woodruffelectronic2001}, interactive museum experiences require visitors to engage in a "sophisticated balancing act", dividing their attention between different information sources. Recent work suggests that the tension inherent in these concerns continue to be a challenge for hybrid design in museums \cite{Back2018-oo}.

Particular challenges apply to house museums: museums that were once houses or homes but which have been transformed into museums displaying and communicating the original interior and functions of the house. Because the house itself is considered a historic artifact, it is challenging to introduce technological installations in such houses because they might conflict with the presentation of an authentic historic interior. Claisse et al. \cite{Claisse2018-vc} formulate four central considerations for the design of experiences for house museums, based on interviews with museum experts:

\begin{enumerate}
\item Maintaining the spirit of the house.
\item Building on the domestic nature of historic houses.
\item Telling stories about, for and by people.
\item Designing for a seamless experience of technology.
\end{enumerate}

Claisse suggests the demands of house museums make them "an interesting and relatively underexplored context for the integration of 'tangible interaction'" \cite{Claisse2016-jb}. Furthermore, Ciolfi and McLoughlin \cite{Ciolfi2011-cw} have formulated some lessons for design of tangible interactions in a museum setting:
\begin{enumerate}
\item Both digital and physical components must fit well within an overall storyline.
\item Tangible artifacts need to be place-sensitive, in order to avoid distracting from the museum setting.
\item The tangible artifacts should be limited to a simple and straightforward functionality, in order to work as "bridging" components between the digital and physical, rather than "high-tech gimmicks".
\end{enumerate}

\section{Approach}
The study at hand was conducted as a Research through Design project \cite{Zimmerman2007jo} from February till December 2018. The process included observations and interviews with visitors at the museum and a co-creation workshop with employees and volunteers at the museum, followed by an iterative design process in which several prototypes were built and tested. Initial prototypes were tested in lab facilities of the university, while the last prototype was integrated and tested in the museum from 27 October to 1 December. The lab tests were done with invited test users as "think aloud" tests with observations and interviews. Data from the observations and interviews were summed up, themed and analyzed in order to form the base of further iterations towards the next prototype. Evaluation of the final prototype was done through observations of the museum's visitors using the prototype in situ.

\section{Context: Greve Museum}
\begin{figure}
\includegraphics[width=\linewidth]{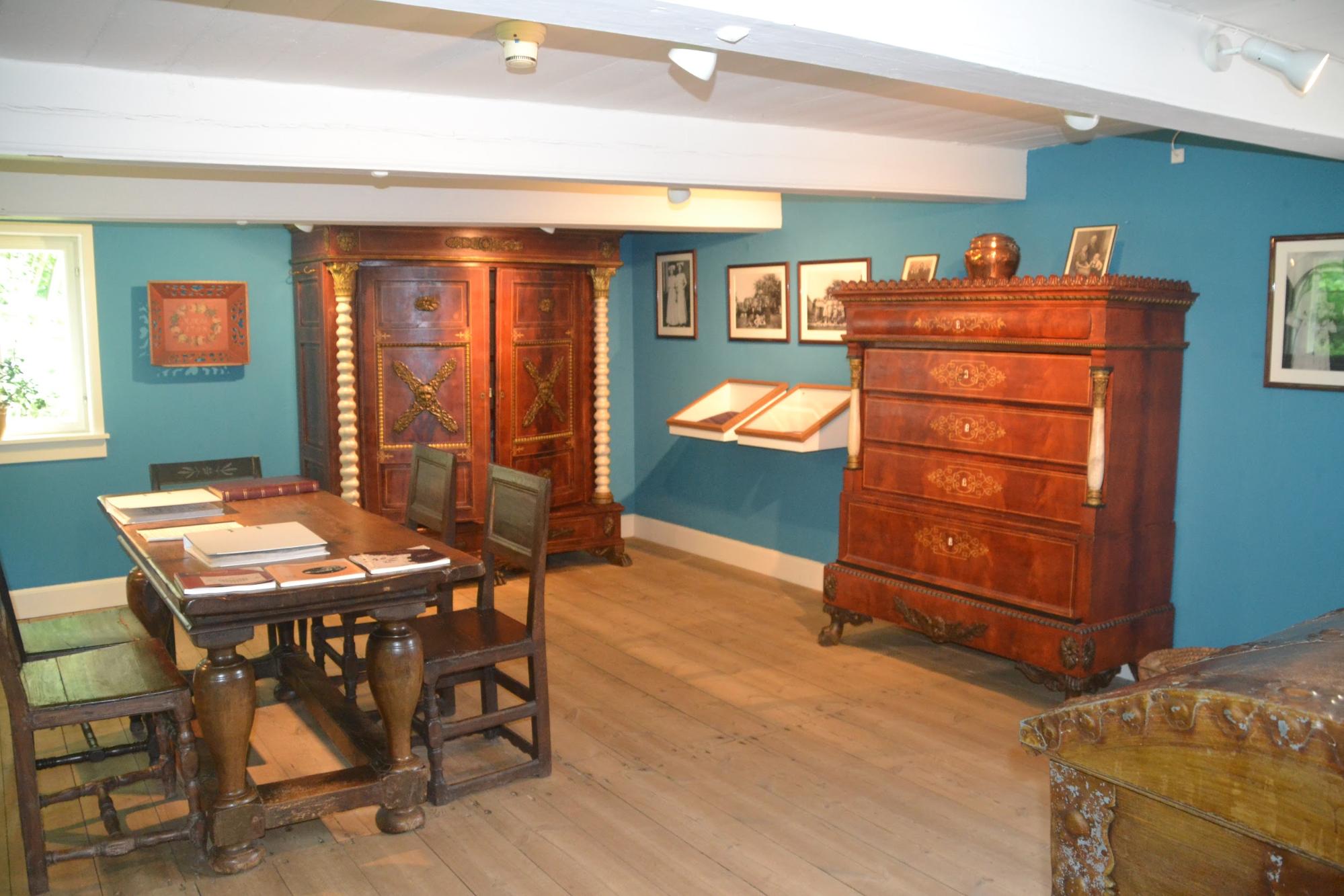}
\caption{\O verstuen before the redesign.}
\label{overstuen}
\end{figure}

Greve Museum is a small local museum exhibiting the historic evolution of Greve village on the outskirts of Copenhagen, Denmark. The museum is located in an old farmhouse furnished and arranged as a farm in this area would be in the 1800s. Our study focuses on a redesign of one of the rooms in the museum, the old storage hall \O verstuen (Fig. \ref{overstuen}). In the 1800s this room had an interesting combination of functions: It was originally used for storage, but was also the room for celebrations as well as for mourning. Being the coldest room on the farm, it was used to lay dead family members on display for mourning, before carrying them out of the "Death Door" - a door specifically used for the dead, as superstition prevented dead bodies from being carried through the door used by the living. However, as museum representatives explained to us at the start of the project, the existing exhibition of the room had failed to engage audiences to any large degree. Guests did not seem to understand this room, or find it interesting enough to spend much time exploring. The museum had decided to redesign the room, removing display cases with old documents and refocusing attention on the room and the artifacts belonging to it, in an attempt to better convey the original functions of the room. As part of this endeavour they asked us to design an interactive experience that could help to better engage the visitors while at the same time providing knowledge about life on the farm in the 1800s. In particular, the museum was hoping to engage interest among younger audiences.

\section{Design}
Our design process started with observations and interviews with visitors in \O everstuen, which confirmed the assessment of the curators: Most visitors barely stepped into the room before turning around and leaving, and few if any engaged with the educational material on display. Subsequently we arranged a co-creation workshop with a group of museum employees and volunteers. Through a structured ideation process we developed a narrative about a ghost haunting the room. Exploring different technologies which could be used to bring this ghost to life, we decided against solutions that would involve smartphones or tablets, as the museum already had a tablet-based experience which was rarely if ever used by visitors. Based on the ideal of seamless integration of technology into the authentic interior we also discarded ideas involving sophisticated display techniques such as augmented reality or Pepper's Ghost.

\begin{figure}
\includegraphics[width=\textwidth]{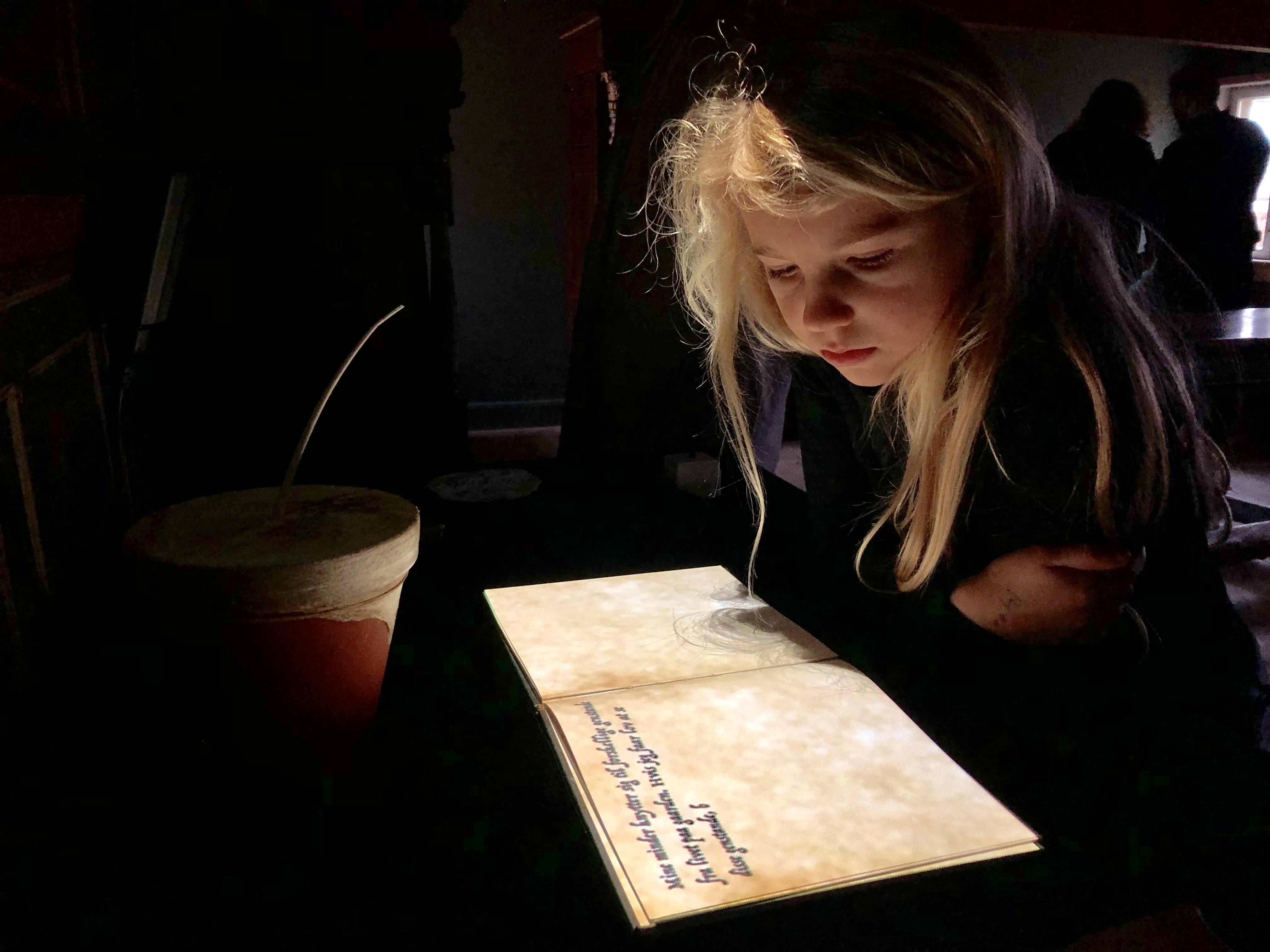}
\caption[]{\textit{The Diary of Niels.}}
\label{diary}
\end{figure}

\begin{figure}
\includegraphics[width=\textwidth]{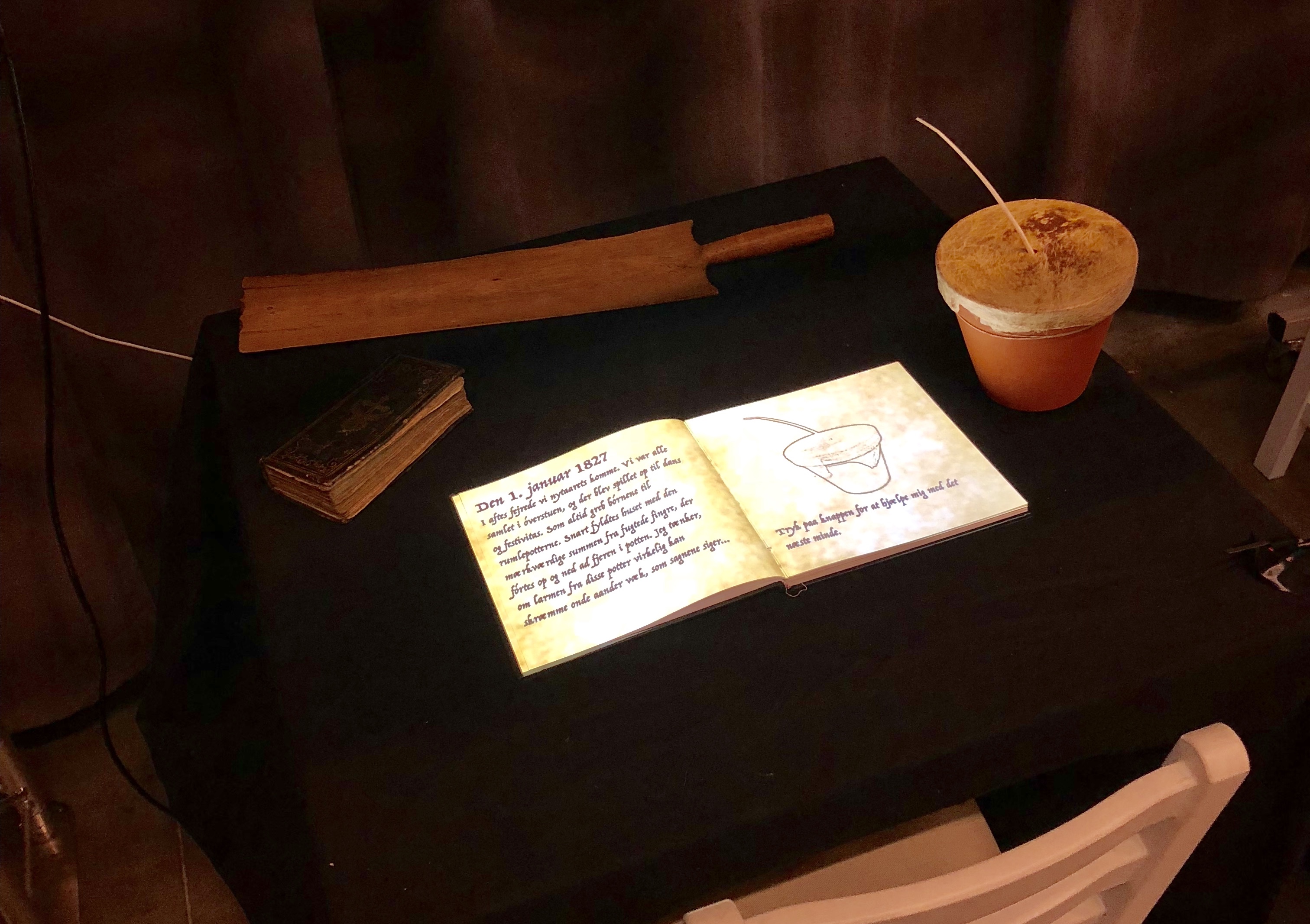}
\caption{The Diary of Niels and the three artifacts used with the diary: A hymn book, an ike beater and a rummelpot.}
\label{artifacts}
\end{figure}

Instead we took inspiration from a scene in a Harry Potter movie \cite{harry_potter}, in which a ghost communicates with Harry by writing on the blank pages of a diary. Through an iterative process we developed a concept consisting of a diary - a physical book with blank pages - lying on a table, with a projector in the ceiling above it (Fig. \ref{diary}). This is the diary of Niels, a historic person who lived at Greve Farm from 1797 to 1870. Writing in the pages of his old diary, the ghost of Niels asks visitors to show him three specific everyday objects from the farm. For each object, Niels responds by sharing a memory from his diary in which this object played a role. This meant combining the physical context of the objects and the room with the digital installation of the diary, thus exploring the opportunities of tangible technology in combination with the potential affective response to the fiction of a ghost in an old farmhouse. Though the diary entries were fictions they were based on the museum's documentation of life on the farm, describing scenarios tied to the everyday objects and how they were used by the farm people. These objects consisted of the following (Fig. \ref{artifacts}): 
\begin{enumerate}
\item A hymnbook which was an important decorational part of the lit de parades in \O everstuen.
\item An ike beater, a wooden tool used by the locals to process fibers when making cloths. 
\item A rummelpot (or friction drum), which was a traditional homemade musical instrument used by farm children on festive occasions. 
\end{enumerate}
While the rummelpot used in the installation was a copy, the hymnbook and ike beater were authentic historical objects, that we were allowed to use due to the fact that they were not registered as archival material in the museum collection.
 
\subsection{Prototyping}
\begin{figure}
\centering
\includegraphics[width=\linewidth]{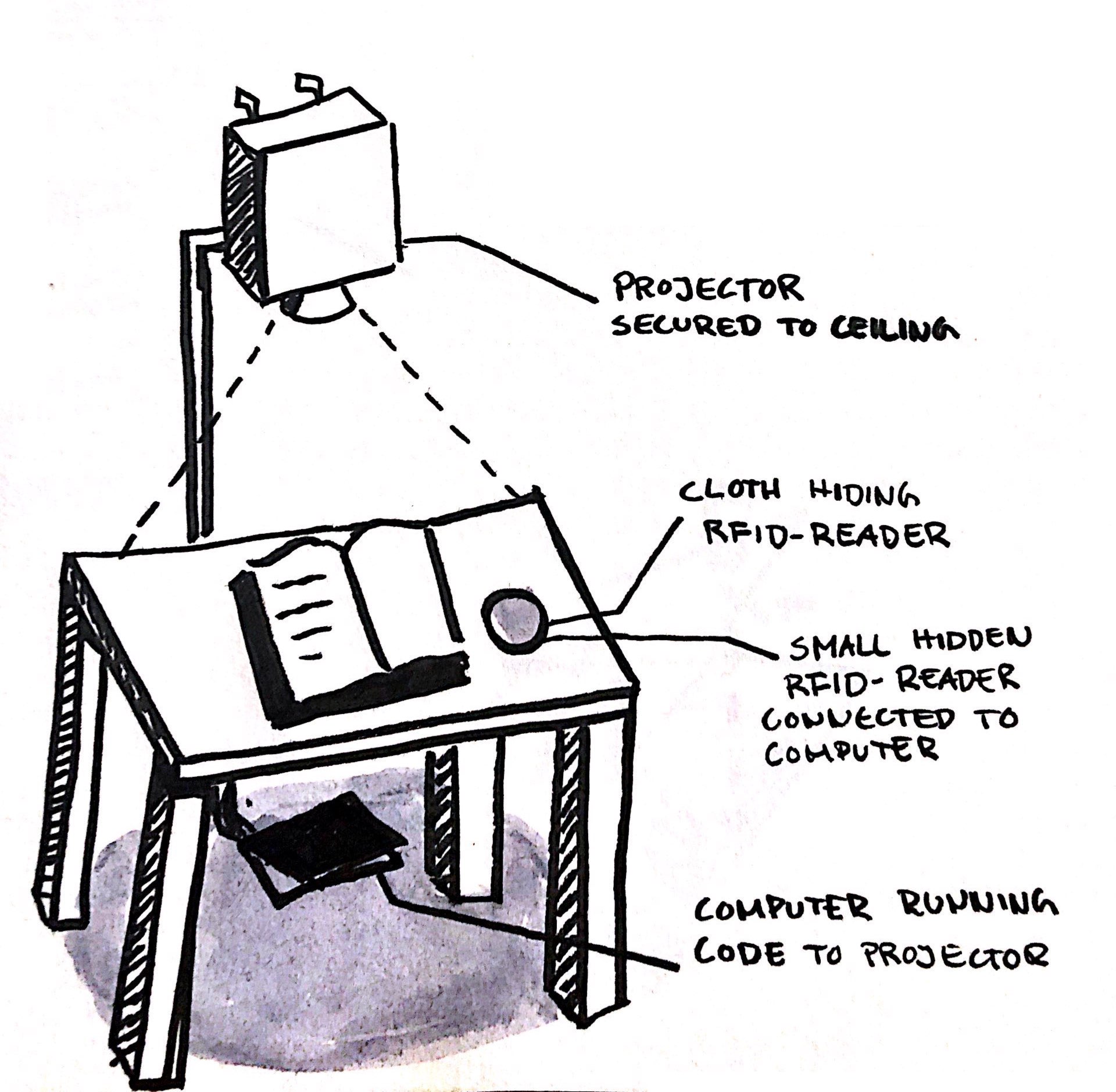}
\caption{Overview of the physical and technical setup of the final prototype.}
\label{sketch}
\end{figure}

To create the sensation of a ghost writing in the diary, the text of the diary was animated and projected down upon the blank pages of a physical book from a projector in the ceiling (Fig. \ref{sketch}). Using a font that looked like handwriting, the letters would appear one at a time, giving the impression that Niels was writing on the pages in real time. The work on our first prototype revolved around building and testing this "magic" diary and projecting it on blank pages. Testing this in a lab setting on two young users (10 and 15 years old), the users found the overall concept entertaining but wanted more variety in interaction forms and questioned the historical authenticity.

In our second prototype we focused on varying the forms of interaction as well as elaborating the historical memories for the diary. Next to the book was now placed a push button which guests were asked to press when they wanted to move forward with the reading. Hidden under a tablecloth, a radio frequency identification (RFID) reader was placed. Three artifacts were each equipped with a set of RFID tags that, when held close to the RFID reader, would trigger a corresponding memory to appear on the pages of the diary. The RFID reader was hidden in order to support the fiction of a supernatural presence in the room. When we tested the second prototype in our lab on three new test users aged 10-11, the test users found the experience engaging and mostly easy to use, in spite of some technical and usability problems. When we asked the users about the contents of the diary, 2 of the 3 were able to recall parts of the content, whereas the third had struggled too much with the readability of the "handwritten" font to remember any of the content. Given that we were aiming to engage children at an age where reading skills are variable, we decided to change the font in order to improve the readability, as well as adjusting the language used in the diary.

\section{Evaluation}
\textit{The Diary of Niels} was implemented in \O everstuen at Greve Museum as a fully functional prototype in November 2018. This "in the wild" deployment gave us the chance to observe how the prototype was used by visitors who had not received any prior instructions or explanations about the prototype. We spent one day in \O everstuen observing how visitors engaged with the installation, interviewing some of them after the experience.

\subsection{Overall reception}
There was a marked difference from our initial observations at \O everstuen before the implementation of the diary. Visitors quickly discovered the diary and started interacting with it, and spent more time exploring the room compared to earlier observations. It should be noted that the entire room had been redesigned in the meantime, so it is impossible to say how much of this improvement was due to \textit{The Diary of Niels}. However, in interviews with the guests they were all able to recall and explain the function of at least one of the three historical objects, indicating that they had engaged sufficiently with the prototype to learn a little about the exhibition. 


\subsection{Group experience}
While the installation had been designed with a single user in mind, visitors actually came to the exhibition and approached the installation in groups (typically families). Thus an adult would do the reading, while the children engaged with the physical objects. This may have helped the educational function of the installation, as the children remembered more about each object than users in the earlier lab test scenarios which had to read themselves. Approaching the diary as families also enabled collective reflection on the exhibition. Visitors discussed the age of the objects, and one visitor approached the adjoining rooms of the farm house, wondering aloud which of the furniture and pictures were Niels' old belongings. This indicates that the diary installation could be used to spark conversations between parents and children, helping them reflect and learn about the exhibition.

\subsection{Place-sensitivity}
Interviews with the visitors revealed a surprising failure of the installation. While the visitors had understood and engaged with much of the content of the diary, recalling events taking place in \O everstuen, they were not aware that they were standing in that very same room in which the events had taken place. This oddity can be explained by the fact that the museum curators had wanted to avoid explanatory texts and signposts on the walls when redesigning the room, so the only text stating the name of the room and explaining its function was an easily overlooked sign posted by the entrance. Meanwhile, while the diary speaks about \O everstuen it does not actually point out that this is the very room the diary is placed in. This illustrates that the design strategy of this installation, aiming to blend in as seamlessly as possible with the historic interior, puts increased demands on the overall exhibition design of the whole room: Curators must consider how much extradiegetic information is needed, and how and where this should be presented to visitors, in order to avoid confusion.

\section{Conclusions}
The setting at Greve museum offered a fertile opportunity for experimenting with tangible interaction, as the museum allowed us to use authentic historic objects in our prototype. This allowed us to use the objects not only as a bridge between the physical and the digital, but also as a material connection between the past and the present. This will not always be possible in other contexts - similar projects have used replicas \cite{marshall_tangible_mesch2016}. However, as many house museums do allow visitors to touch and pick up many of their artifacts, there may be possibilities to explore tangible interactions with authentic artifacts by using non-invasive techniques, e.g. visual object recognition.

The problem we encountered with the place-sensitivity of our prototype points to a need for designers to look holistically at the information provided to visitors, not just through the interactive system but also the rest of the room and the museum as a whole. In our design process we had explored the possibility of integrating the installation more extensively with the rest of the room by turning it into a scavenger hunt, in which the objects belonging to the diary installation would be hidden around the room. This would have required users to search and explore the room in order to unlock the diary entries, thus extending the interaction into the entire room rather than just the tabletop. However, this idea had to be discarded due to limitations set by the museum. Further effort could be made to connect the diary with the room, for instance by installing other personal traces of Niels in the room so that the narrative of the diary would be conveyed through several different means in different parts of the room. These traces could e.g. be "micro-augmentations" \cite{Antoniou2015} such as sound installations of Niels whispering when a visitor approached certain objects. Future research should explore opportunities for further integrating artifact interactions with the experience of the rest of the museum space. 

\section{Acknowledgments}
The project has received funding from the European Union’s Horizon 2020 research and innovation programme under grant agreement No 727040 (the GIFT project: gifting.digital).

%
%
%

\bibliographystyle{splncs04}
\bibliography{main}

\begin{thebibliography}{10}
\providecommand{\url}[1]{\texttt{#1}}
\providecommand{\urlprefix}{URL }
\providecommand{\doi}[1]{https://doi.org/#1}

\bibitem{Antoniou2015}
Antoniou, A., O'Brien, J., Bardon, T., Barnes, A., Virk, D.:
  Micro-augmentations: Situated calibration of a novel non-tactile, peripheral
  museum technology. In: Proceedings of the 19th Panhellenic Conference on
  Informatics. pp. 229--234. PCI '15, ACM, New York, NY, USA (2015).
  \doi{10.1145/2801948.2801959},
  \url{http://doi.acm.org/10.1145/2801948.2801959}

\bibitem{Back2018-oo}
Back, J., Bedwell, B., Benford, S., Eklund, L., L{\o}vlie, A.S., Preston, W.,
  Rajkowska, P., Ryding, K., Spence, J., Thorn, E.C., Waern, A., Wray, T.:
  {{GIFT}}: {Hybrid} {Museum} {Experiences} through {Gifting} and {Play}. In:
  Antoniou, A., Wallace, M. (eds.) Proceedings of the {Workshop} on {Cultural}
  {Informatics} co-located with the {EUROMED} {International} {Conference} on
  {Digital} {Heritage} 2018 ({{EUROMED}} 2018). vol.~2235, pp. 31--40. CEUR
  Workshop Proceedings, Nicosia, Cyprus (2018)

\bibitem{Bannon2005-mj}
Bannon, L., Benford, S., Bowers, J., Heath, C.: Hybrid design creates
  innovative museum experiences. Commun. ACM  \textbf{48}(3),  62--65 (Mar
  2005)

\bibitem{Blackman2012-hc}
Blackman, L.: Immaterial Bodies: Affect, Embodiment, Mediation. SAGE, Newbury
  Park, CA (2012)

\bibitem{Boehner_depaula_2005}
Boehner, K., DePaula, R., Dourish, P., Sengers, P.: Affect: From information to
  interaction. In: Proceedings of the 4th Decennial Conference on Critical
  Computing: Between Sense and Sensibility. pp. 59--68. CC '05, ACM, New York,
  NY, USA (2005)

\bibitem{Boehner_sengers_2005}
Boehner, K., Sengers, P., Gay, G.: Affective presence in museums: Ambient
  systems for creative expression. Digital Creativity  \textbf{16}(2),  79--89
  (Jan 2005)

\bibitem{Ciolfi2011-cw}
Ciolfi, L., McLoughlin, M.: Physical keys to digital memories: reflecting on
  the role of tangible artefacts in ``reminisce''. p.~10 (2011)

\bibitem{Claisse2016-jb}
Claisse, C.: Crafting tangible interaction to prompt visitors' engagement in
  house museums. In: Proceedings of the {TEI} '16: Tenth International
  Conference on Tangible, Embedded, and Embodied Interaction. pp. 681--684. ACM
  (Feb 2016)

\bibitem{Claisse2018-vc}
Claisse, C., Petrelli, D., Dulake, N., Marshall, M., Ciolfi, L.: Multisensory
  interactive storytelling to augment the visit of a historical house museum.
  In: Proceedings of the 2018 Digital Heritage International Congress. IEEE
  (Aug 2018)

\bibitem{harry_potter}
Columbus, C.: Harry potter and the chamber of secrets (2002)

\bibitem{Gregory2007-vi}
Gregory, K., Witcomb, A.: Beyond nostalgia: the role of affect in generating
  historical understanding at heritage sites. Museum revolutions: how museums
  change and are changed pp. 263--275 (2007)

\bibitem{Hornecker2006-zn}
Hornecker, E., Buur, J.: Getting a grip on tangible interaction: A framework on
  physical space and social interaction. In: Proceedings of the {SIGCHI}
  Conference on Human Factors in Computing Systems. pp. 437--446. CHI '06, ACM,
  New York, NY, USA (2006)

\bibitem{vom_lehn}
vom Lehn, D., Heath, C.: Displacing the {Object}: {Mobile} {Technologies} and
  {Interpretive} {Resources}. In: International {Cultural} {Heritage}
  {Informatics} {Meeting}: {Proceedings} from ichim03. Archives \& Museum
  Informatics, Paris (2003),
  \url{http://www.archimuse.com/publishing/ichim03/088C.pdf}

\bibitem{lovlie_gift_2019}
Løvlie, A.S., Benford, S., Spence, J., Wray, T., Mortensen, C.H., Olesen, A.,
  Rogberg, L., Bedwell, B., Darzentas, D., Waern, A.: The {GIFT} framework:
  {Give} visitors the tools to tell their own stories. In: Museums and the
  {Web} 2019. MW18: Museums and the Web, Boston, MA, USA (Apr 2019),
  \url{https://mw19.mwconf.org/paper/the-gift-framework-give-visitors-the-tools-to-tell-their-own-stories/}

\bibitem{lovlie_designing}
Løvlie, A.S., Eklund, L., Waern, A., Ryding, K., Rajkowska, P.: Designing for
  interpersonal museum experiences. In: Black, G. (ed.) Museums and the
  {Challenge} of {Change}: old institutions in a new world. Routledge, London
  \& New York (in press)

\bibitem{marshall_tangible_mesch2016}
Marshall, M.T., Dulake, N., Ciolfi, L., Duranti, D., Kockelkorn, H., Petrelli,
  D.: Using tangible smart replicas as controls for an interactive museum
  exhibition. In: Proceedings of the TEI '16: Tenth International Conference on
  Tangible, Embedded, and Embodied Interaction. pp. 159--167. TEI '16, ACM, New
  York, NY, USA (2016). \doi{10.1145/2839462.2839493},
  \url{http://doi.acm.org/10.1145/2839462.2839493}

\bibitem{Meskin2003-rq}
Meskin, A., Weinberg, J.M.: Emotions, fiction, and cognitive architecture. Brit
  J Aesthetics  \textbf{43}(1),  18--34 (Jan 2003)

\bibitem{pedersendesigning2019}
Pedersen, T., Andersen, E.T., Løvlie, A.S.: Designing a “{No} {Interface}”
  {Audio} {Walk}. In: Museums and the {Web} 2019. Museums and the Web, Boston,
  MA, USA (Apr 2019),
  \url{https://mw19.mwconf.org/paper/designing-a-no-interface-audio-walk/}

\bibitem{Petrelli2018-bd}
Petrelli, D., O'Brien, S.: Phone vs. tangible in museums: A comparative study.
  In: Proceedings of the 2018 {CHI} Conference on Human Factors in Computing
  Systems. pp. 112:1--112:12. CHI '18, ACM, New York, NY, USA (2018)

\bibitem{Prown1980-pk}
Prown, J.D.: Style as evidence. Winterthur Portf.  \textbf{15}(3),  197--210
  (Oct 1980)

\bibitem{risseeuw_authoring_2016}
Risseeuw, M., Cavada, D., Not, E., Zancanaro, M., Marshall, M.T., Petrelli, D.,
  Kubitza, T.: Authoring {Augmented} {Digital} {Experiences} in {Museums}. In:
  Proceedings of the {International} {Working} {Conference} on {Advanced}
  {Visual} {Interfaces}. pp. 340--341. {AVI} '16, ACM, New York, NY, USA
  (2016). \doi{10.1145/2909132.2926064},
  \url{http://doi.acm.org/10.1145/2909132.2926064}

\bibitem{Rudloff2013-oi}
Rudloff, M.: Det medialiserede museum: digitale teknologiers transformation af
  museernes formidling. MedieKultur: Journal of media and communication
  research  \textbf{29}(54), ~22 (2013)

\bibitem{Ryding_Fritsch}
Ryding, K., Fritsch, J.: Play design as a relational strategy to intensify
  affective encounters in the art museum. In: Proceedings of the 2020 ACM
  Designing Interactive Systems Conference. p. 681–693. DIS '20, Association
  for Computing Machinery, New York, NY, USA (2020).
  \doi{10.1145/3357236.3395431}, \url{https://doi.org/10.1145/3357236.3395431}

\bibitem{Smith2018-ge}
Smith, L., Wetherell, M., Campbell, G.: Emotion, Affective Practices, and the
  Past in the Present. Routledge, London (Jun 2018)

\bibitem{wessel_potentials_2007}
Wessel, D., Mayr, E.: Potentials and {Challenges} of {Mobile} {Media} in
  {Museums}. International Journal of Interactive Mobile Technologies (iJIM)
  \textbf{1}(1) (Oct 2007),
  \url{http://journals.sfu.ca/onlinejour/index.php/i-jim/article/view/165}

\bibitem{witcomb2013}
Witcomb, A.: Understanding the role of affect in producing a critical pedagogy
  for history museums. Museum Management and Curatorship  \textbf{28}(3),
  255--271 (2013). \doi{10.1080/09647775.2013.807998},
  \url{https://doi.org/10.1080/09647775.2013.807998}

\bibitem{Witcomb2014-wq}
Witcomb, A.: ``look, listen and feel'': The first peoples exhibition at the
  bunjilaka gallery, melbourne museum. Thema La revue des Mus{\'e}es de la
  civilistion  \textbf{1},  49--62 (Jan 2014)

\bibitem{woodruffelectronic2001}
Woodruff, A., Aoki, P., Hurst, A., Szymanski, M.: Electronic {Guidebooks} and
  {Visitor} {Attention}. In: Bearman, D., Garzotto, F. (eds.) International
  Cultural Heritage Informatics Meeting: Proceedings from ichim01. Archives \&
  Museum Informatics, Milano, Italy (2001),
  \url{http://www.archimuse.com/publishing/ichim01\_vol1/woodruff.pdf}

\bibitem{zancanarorecipes2015}
Zancanaro, M., Not, E., Petrelli, D., Marshall, M., van Dijk, T., Risseeuw, M.,
  van Dijk, D., Venturini, A., Cavada, D., Kubitza, T.: Recipes for tangible
  and embodied visit experiences. In: {MW}2015: {Museums} and the {Web} 2015.
  Museums and the Web, Chicago, IL (2015),
  \url{https://mw2015.museumsandtheweb.com/paper/recipes-for-tangible-and-embodied-visit-experiences/}

\bibitem{Zimmerman2007jo}
Zimmerman, J., Forlizzi, J., Evenson, S.: Research through design as a method
  for interaction design research in {HCI}. In: Proceedings of the {SIGCHI}
  Conference on Human Factors in Computing Systems. pp. 493--502. ACM (Apr
  2007)

\end{thebibliography}

\end{document}